\documentclass[10pt]{IEEEtran}
\ifCLASSINFOpdf
\else
\fi
\usepackage{amssymb}
\usepackage{multirow}
\usepackage{indentfirst}
\usepackage{inputenc}
\usepackage{array}
\usepackage[switch,mathlines]{lineno}
\usepackage{soul}
\usepackage{graphicx}
\usepackage{cite}
\usepackage[cmex10]{amsmath}
\usepackage{algorithmic}
\usepackage{array}
\usepackage{fixltx2e}
\usepackage{CJK}
\usepackage[cmex10]{amsmath}
\usepackage{algorithmic}
\usepackage{array}
\usepackage{fixltx2e}
\usepackage{bm}
\usepackage{color}
\usepackage{amsmath,amsthm,amssymb,amsfonts}
\newtheorem{theorem}{Theorem}

\newtheorem*{remark}{Remark}
\newtheorem{proposition}[theorem]{Proposition}
\usepackage{arydshln} 
\usepackage{soul}
\usepackage{tikz}
\usepackage[caption=false,font=footnotesize]{subfig}
\usepackage{fixltx2e}
\begin{document}
%
\title{Robust Power System Dynamic State Estimator with Non-Gaussian Measurement Noise: Part II--Implementation and Results}
\author{Junbo~Zhao,~\IEEEmembership{Student Member,~IEEE},~Lamine~Mili,~\IEEEmembership{Fellow,~IEEE}\\

\thanks{Junbo Zhao and Lamine Mili are with the Bradley Department of Electrical Computer Engineering, Virginia Polytechnic Institute and State University, Northern Virginia Center, Falls Church, VA 22043, USA (e-mail: zjunbo@vt.edu, lmili@vt.edu).}}

\markboth{IEEE TRANSACTIONS ON POWER SYSTEMS,~Vol.~, No.~, ~2017}%
{Shell \MakeLowercase{\textit{et al.}}: Bare Demo of IEEEtran.cls for Journals}
\maketitle

\begin{abstract}
This paper is the second of a two-part series that discusses the implementation issues and test results of a robust Unscented Kalman Filter (UKF) for power system dynamic state estimation with non-Gaussian synchrophasor measurement noise. The tuning of the parameters of our Generalized Maximum-Likelihood-type robust UKF (GM-UKF) is presented and discussed in a systematic way. Using simulations carried out on the IEEE 39-bus system, its performance is evaluated under different scenarios, including i) the occurrence of two different types of noises following thick-tailed distributions, namely the Laplace or Cauchy probability distributions for real and reactive power measurements; ii) the occurrence of observation and innovation outliers; iii) the occurrence of PMU measurement losses due to communication failures; iv) cyber attacks; and v) strong system nonlinearities. It is also compared to the UKF and the Generalized Maximum-Likelihood-type robust iterated EKF (GM-IEKF). Simulation results reveal that the GM-UKF outperforms the GM-IEKF and the UKF in all scenarios considered. In particular, when the system is operating under stressed conditions, inducing system nonlinearities, the GM-IEKF and the UKF diverge while our GM-UKF does converge. In addition, when the power measurement noises obey a Cauchy distribution, our GM-UKF converges to a state estimate vector that exhibits a much higher statistical efficiency than that of the GM-IEKF; by contrast, the UKF fails to converge. Finally, potential applications and future work of the proposed GM-UKF are discussed in concluding remarks section.
\end{abstract}
\vspace{-0.15cm}

%
\IEEEpeerreviewmaketitle
\begin{figure*}[!t]
\centering
\includegraphics[height=6.5cm]{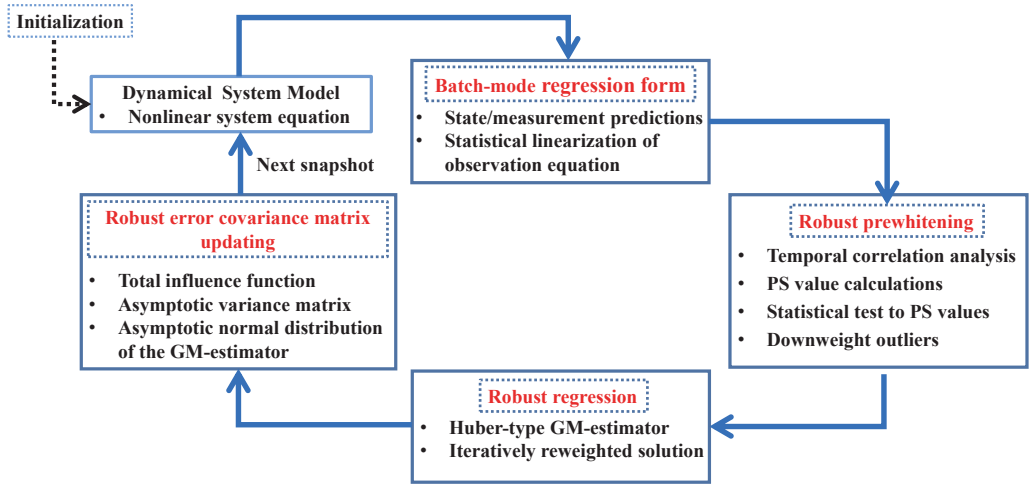}
\caption{Flowchart representing the four main steps of the proposed GM-UKF.}
\label{Fig.flowchart_algorithm}
\end{figure*}
\vspace{-0.4cm}
\section{Introduction}
\IEEEPARstart{R}{eliable} and fast dynamic state estimator (DSE) plays a vital role in power system monitoring and control. In the literature, both the process and the observation noises of the system nonlinear dynamic models are assumed to be Gaussian when developing a DSE. Furthermore, the dynamical system model is supposed to be accurate and the PMU measurements are secure. However, these assumptions do not hold true for practical power systems as elaborated in the first part of this two-part series. To address these problems, several robust Extended Kalman Filter (EKF) and Unscented Kalman Filter (UKF)-based DSEs have been proposed \cite{Rouhani2016,Wang2010,Chang2012,Junbo_GMIEKF2016,Junbo_PESGM2017}. In  \cite{Rouhani2016,Wang2010,Chang2012}, while filters based on the Huber M-estimator are proposed to suppress observation outliers, they are vulnerable to measurement noise obeying thick-tailed distribution and innovation outliers that are induced by model parameter errors. These limitations are mitigated in \cite{Junbo_GMIEKF2016,Junbo_PESGM2017} via a Generalized Maximum-Likelihood-type robust iterated EKF (GM-IEKF). While the latter is resistant to observation and innovation outliers and several types of cyber attacks, it has poor statistical efficiency under thick-tailed non-Gaussian measurement noises. In addition, due to the inherent limitation of the EKF, the GM-IEKF may fail to converge when the system is operating under stressed conditions, that is, when exhibiting strong model nonlinearities.

In Part I \cite{JunboMili2016}, we develop a Generalized Maximum-Likelihood-type robust UKF-based DSE, termed the GM-UKF for short. We show that our GM-UKF is able to handle thick-tailed non-Gaussian measurement noises with good statistical efficiency and is robust to observation and innovation outliers. In this second part, we will focus on its implementation. Specifically, we will discuss how to set and tune the parameters of the GM-UKF along with the choice of state initialization of the algorithm that solve for the filter. We will then evaluate the performance of our GM-UKF through the following scenarios: i) the occurrence of two different types of noises following Laplace or Cauchy probability distributions for the real and reactive power measurements; ii) the occurrence of observation and innovation outliers; iii) the occurrence of PMU measurement losses due to occasional communication failures; iv) cyber attacks; and v) strong system nonlinearities. WE perform comparisons between our GM-UKF and the UKF and the GM-IEKF proposed in \cite{Junbo_GMIEKF2016}. We show that our GM-UKF outperforms the GM-IEKF and the UKF for all the scenarios being considered. When the system is operating under stressed conditions, our GM-UKF converges while the GM-IEKF and the UKF diverge. Furthermore, if the power measurement noises follow a Cauchy distribution, the UKF fails to converge; although the GM-IEKF can handle that case, it has a much lower statistical relative efficiency with respect to the GM-UKF.

In this paper, the statistical (asymptotic) efficiency of an estimator at a given probability distribution (e.g., Gaussian, or Laplacian, or Cauchy distribution) is defined as the ratio between the inverse of the Fisher information evaluated at that distribution and the (asymptotic) variance of the normalized estimator when all the assumptions underlying the model are exact. As for the robustness of an estimator to outliers, Hampel \cite{Hampel1986} proposes to investigate how the asymptotic bias and the asymptotic variance of the estimator increase with the fraction of contamination, $\epsilon$, termed bias- and variance-robustness analysis. To quantify bias-robustness, he introduces the (bias-)breakdown point, the influence function and the asymptotic maximum bias curve while to quantify variance-robustness, he introduces the change-of-variance function, which measures the sensitivity of the asymptotic variance to an infinitesimal change in $\epsilon$ about zero. A robust estimator has a finite bias and a finite variance when subject to contamination up to the breakdown point. Using simulations carried out on the IEEE 39-bus system, we will show that our GM-UKF satisfies all these robustness and efficiency properties.

The paper is organized as follows. Section II deals with the modeling of power system dynamics, implementation issues and a systematic way to tune parameters. Section III shows the test results under several scenarios using the detailed two-axis generator models. Finally, Section IV concludes the paper and presents some interesting future research directions.
\vspace{-0.4cm}
\section{Algorithm Implementation}
The developed robust GM-UKF is a generic technique for online monitoring of many dynamical cyber-physical systems, including smart grids, autonomous vehicles, aircraft tracking, GPS tracking and navigation, radar systems, to name a few. In this paper, we take power system dynamic state estimation as an illustrative example to demonstrate the capabilities of our GM-UKF for suppressing observation and innovation outliers while being able to filter out thick-tailed non-Gaussian noise.
\vspace{-0.6cm}
\subsection{Nonlinear Discrete-Time Power System Dynamical Model}
For an electric power system, its discrete-time state space representation can be expressed as
\begin{equation}
{\bm{x}_k} = \bm{f}\left( {{\bm{x}_{k - 1}},{{\bm{u}}_{k}}} \right) + {{\bm{w}}_{k}},
\label{Eq:discrete_PSstate_model}
\end{equation}
\begin{equation}
\bm{z}_k = {\bm{h}}\left( {{\bm{x}_k},{{\bm{u}}_k}} \right) + {{\bm{v}}_k},
\label{Eq:discrete_PSobservation_model}
\end{equation}
where the state vector ${\bm{x}_k}$ contains the rotor angle, the rotor speed, the d- and q- axis state variables of the synchronous generator, the exciter, the voltage regulator, and the governor. Here, $\bm{u}_k$ represents the input vector; $\bm{h}(\cdot)$ is the vector-valued measurement function while $\bm{f}(\cdot)$ is the vector-valued function that relates $\bm{x}_k$ to $\bm{x}_{k-1}$; $\bm{z}_k $ is the measurement vector that contains a collection of voltage phasors, current phasors, real and reactive power flows and power injections so that the system dynamical model is observable. The noises $\bm{w}_k$ and $\bm{v}_k$, which may be non-Gaussian noise, are assumed to be white and independent of each other. In this paper, the detailed 9th order two-axis generator model with IEEE-DC1A exciter and TGOV1 turbine-governor is assumed and tested \cite{PES2007}.
\vspace{-0.4cm}
\subsection{Implementation of the GM-UKF}
The flowchart of the proposed GM-UKF is shown in Fig. \ref{Fig.flowchart_algorithm}. It consists of four major steps, namely a batch-mode regression form step, a robust pre-whitening step, a robust regression and robust error covariance matrix updating steps. Specifically, after state initialization and the application of statistical linearization to the nonlinear system process model, we calculate the predicted state and its associated covariance matrix. Next, applying statistical linearization to the nonlinear observation function around the predicted state, we derive the expression of the predicted measurement and its covariance matrix. Then, by processing the observations and predictions simultaneously, we obtain the batch-mode regression form. Next, we apply the PS to a matrix that consists of two-time sequence of the predicted state and innovation vectors to detect the presence of any observation and innovation outliers. This in turn allows us to carry out a robust prewhitening of the regression model. To suppress outliers and filter out thick-tailed non-Gaussian measurement noise, the GM-estimator is used and solved by means of the Iteratively Reweighted Least Squares (IRLS) algorithm. Finally, the total influence function of our GM-UKF is derived and utilized to derive the asymptotic state estimation error covariance matrix. Note that during the iterative solution of the GM-UKF, we advocate to use the Weighted Least Squares (WLS) estimation for the first iteration and then switch to the IRLS algorithm. By doing so we improve the convergence speed of that algorithm.
\vspace{-0.2cm}
\begin{remark}
Corollary 3.1 in Part I of the two-part series states that the matrix $\bm{Z}$ roughly follows a bivariate Gaussian distribution. Using that property, we carry out extensive Monte Carlo simulations and QQ-plots to determine the probability distribution of the PS. From Fig. 3 in Part I, we infer that they approximately obey a chi-squares distribution with 2-degrees of freedom. Consequently, we set the outlier detection threshold of the statistical test applied to the PS to $\eta=\chi_{2,0.975}^2$ at a significance level of 97.5\%. The detailed implementation procedures of the PS algorithm are presented in Appendix A.
\end{remark}
\vspace{-0.5cm}
\subsection{Tuning the Parameters of the GM-UKF}
The tuning of the GM-UKF involves the settings of the breakpoint $\lambda$ of the Huber $\rho$-function, the parameter $d$ of the weighting function, and the convergence tolerance of the IRLS algorithm. $\lambda$ determines the trade-off that we wish to achieve between a least-squares and a least-absolute-value fit. Indeed, when $\lambda\rightarrow 0$, the Huber $\rho$-function tends to the least-absolute-value $\rho$-function and when $\lambda\rightarrow \infty$, it tends to the least-squares $\rho$-function. Regarding the parameter $d$, it determines the statistical efficiency of the PS at the assumed probability distribution along with the robustness of the GM-estimator \cite{Lmili2010}. Decreasing this parameter too much shrinks the dimensions of the 97.5\% confidence ellipse. As a result, good measurements may be unduly downweighted, which yields a decrease in the statistical efficiency. On the other hand, increasing $d$ will increase the bias of the GM-estimator. Extensive simulations have shown that the parameters $\lambda$ and $d$ can be set to 1.5 to achieve a good statistical efficiency at the Gaussian, the Laplacian, and the Cauchy distributions while achieving a good robustness to outliers. Regarding the convergence tolerance threshold of the IRLS algorithm, a typical value is $0.01$; decreasing this value results in small incremental changes of the state estimates while increasing the computing time of the algorithm.
\vspace{-0.1cm}
\begin{proposition}
The coefficient of the estimation error covariance matrix of our GM-UKF expressed as $\frac{{{\mathbb{E}_\Phi }\left[ {{\psi ^2}\left( {{r_{{S}}}} \right)} \right]}}{{{{\left\{ {{\mathbb{E}_\Phi}\left[ {{\psi ^{\prime}}\left( {{r_{{S}}}} \right)} \right]} \right\}}^2}}}$ is equal to 1.0369 for the Huber cost function with $\lambda$=1.5.
\end{proposition}
\vspace{-0.1cm}
\begin{proof}
See the proof in Appendix B.
\end{proof}

\vspace{-0.3cm}
\section{Numerical Results}
The performances of our GM-UKF to handle thick-tailed non-Gaussian noise along with innovation and observation outliers are assessed on the IEEE 39-bus system. The UKF and the GM-IEKF \cite{Junbo_GMIEKF2016} are included for comparisons. The time-domain simulation results are used to generate a collection of samples of the nodal voltage magnitudes and phase angles as well as of the real and reactive power injections at the terminal buses of all the generators. A sampling rate of 50 samples/second is assumed. A synthetic noise is added to the true values following the probability distributions displayed in Fig. 1 of Part I. Specifically, a zero mean Gaussian noise is assumed for the voltage angles, a bimodal Gaussian mixture distribution is assumed for the noise of the voltage magnitudes, and either a Laplace or a Cauchy distribution is assumed for the noise of the real and reactive power measurements. Note that the random variable that follows a Gaussian mixture distribution is generated via Matlab functions; the Cauchy random variable $\varrho$ is obtained by sampling the inverse cumulative distribution function of the distribution given by
\begin{figure}
\centering
  \mbox{\subfloat[]{\label{subfig:a} \includegraphics[width=9.3cm]{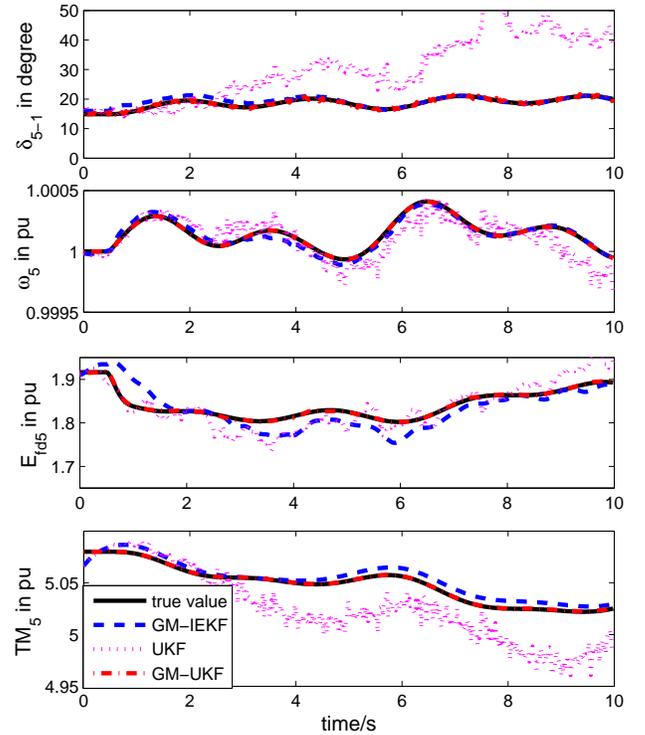}}}
  \mbox{\subfloat[]{\label{subfig:b} \includegraphics[width=7cm]{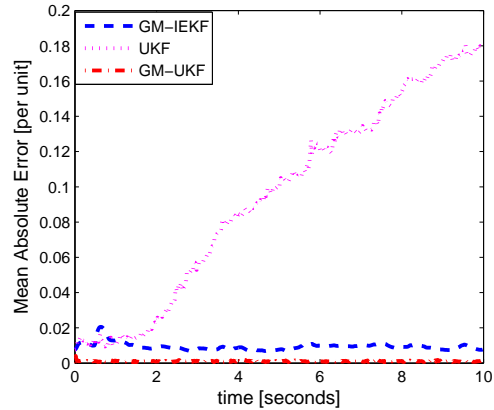}}}
\caption{Case 1: Tracking performance of the GM-IEKF, UKF, and GM-UKF without outliers; (a) the estimated rotor angle and speed, field voltage and mechanical power of Generator 5 are used for illustration purposes; (b) mean absolute error of each of the three filters.}
\label{Fig.Laplace_without_outliers}
\end{figure}
\vspace{-0.1cm}
\begin{equation}
\varrho = \beta+\alpha \cdot \tan \left( {\pi \left( {U_1 - 0.5} \right)} \right),
\end{equation}
where $\beta$ is the location and $\alpha$ is the scale parameter; $U_1$ are values randomly sampled from the uniform distribution on the interval (0,1); samples obeying the Laplace random variable $\zeta$ with mean $\mu$ and scale $b$ are generated using
\begin{equation}
\zeta = \mu  - b\;{\mathop{\rm sgn}} \left( U_2 \right)\ln \left( {1 - 2\left| U_2 \right|} \right),
\end{equation}
where $U_2$ is a random variable drawn from the uniform distribution in the interval (1/2, 1/2].

The two-axis generator model is assumed and tested, whose parameters are taken from \cite{PES2007}. A disturbance is applied at $t$=0.5s by opening the transmission line between Buses 15 and 16. The maximal number of iterations allowed for the IRLS algorithm is 20. For the state initialization, the steady-state values with 10\% errors are used. Due to space limitation, not all the 9 state variables of each generator are shown; instead estimated values of the rotor angle and speed, the field voltage and the mechanical power of Generator 5 are utilized for illustration purposes. The mean absolute error (MAE) is utilized as the index to evaluate the overall performance of each method.

\subsection{Case 1: Thick-tailed Non-Gaussian Measurement Noise without Outliers}
In this section, we first evaluate the performance of the GM-IEKF, the UKF, and the GM-UKF under normal conditions. Specifically, a zero mean Gaussian noise with a standard deviation of $10^{-2}$ is added to the voltage angles; the noise of the voltage magnitudes follows a bimodal Gaussian mixture with zero mean, variances of $10^{-4}$ and $10^{-3}$ and weights of 0.9 and 0.1, respectively; Laplace noise with zero mean and scale 0.2 is added to the real and reactive power injections measurements. The test results are displayed in Fig. \ref{Fig.Laplace_without_outliers}. It is observed that the UKF is not able to cope with Laplacian noise even in the absence of outliers. By contrast, the GM-IEKF and the GM-UKF can filter out such a noise while achieving good tracking performance. However, the GM-IEKF has much lower relative statistical efficiency with respect to our GM-UKF. In particular, the GM-IEKF poorly estimates the field voltage and the rotor speed. By observing Fig. \ref{Fig.Laplace_without_outliers}, it is interesting to note that the turbine mechanical power is changing during the transient process. According to the CIGRE report \cite{CIGRE_report2003}, it can significantly vary when control features such as fast valving or special protection schemes are used to limit the output of the steam driven generator during transients. Consequently, it is of vital importance to not assume it to be fixed at a constant steady-state value as commonly done in most of the literature, but to obtain accurate dynamic state variables of the governor for controls and stability analysis.
\vspace{-0.3cm}

\begin{figure}
\centering
  \mbox{\subfloat[]{\label{subfig:a} \includegraphics[width=9.5cm]{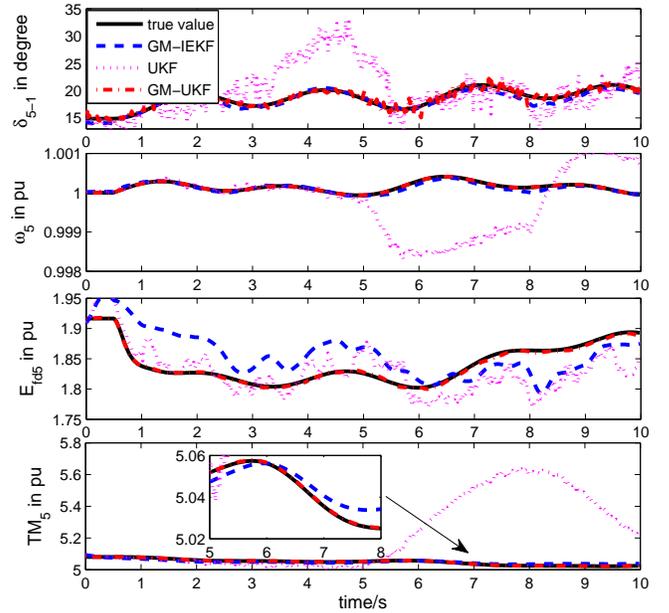}}}
  \mbox{\subfloat[]{\label{subfig:b} \includegraphics[width=7cm]{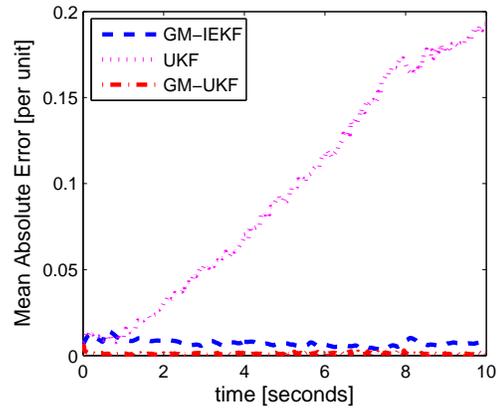}}}
\caption{Case 2: Tracking performance of the GM-IEKF, the UKF, and the GM-UKF in the presence of observation outliers from $t$=4s to $t$=6s, where the real and reactive power measurements of Generator 5 are corrupted with 20\% errors; (a) the estimated rotor angle and speed, field voltage and mechanical power of Generator 5 are used for illustration purposes; (b) mean absolute error of each of the three filters.}
\label{Fig.Laplace_observation_outliers}
\end{figure}

\begin{figure}
\centering
  \mbox{\subfloat[]{\label{subfig:a} \includegraphics[width=9.5cm]{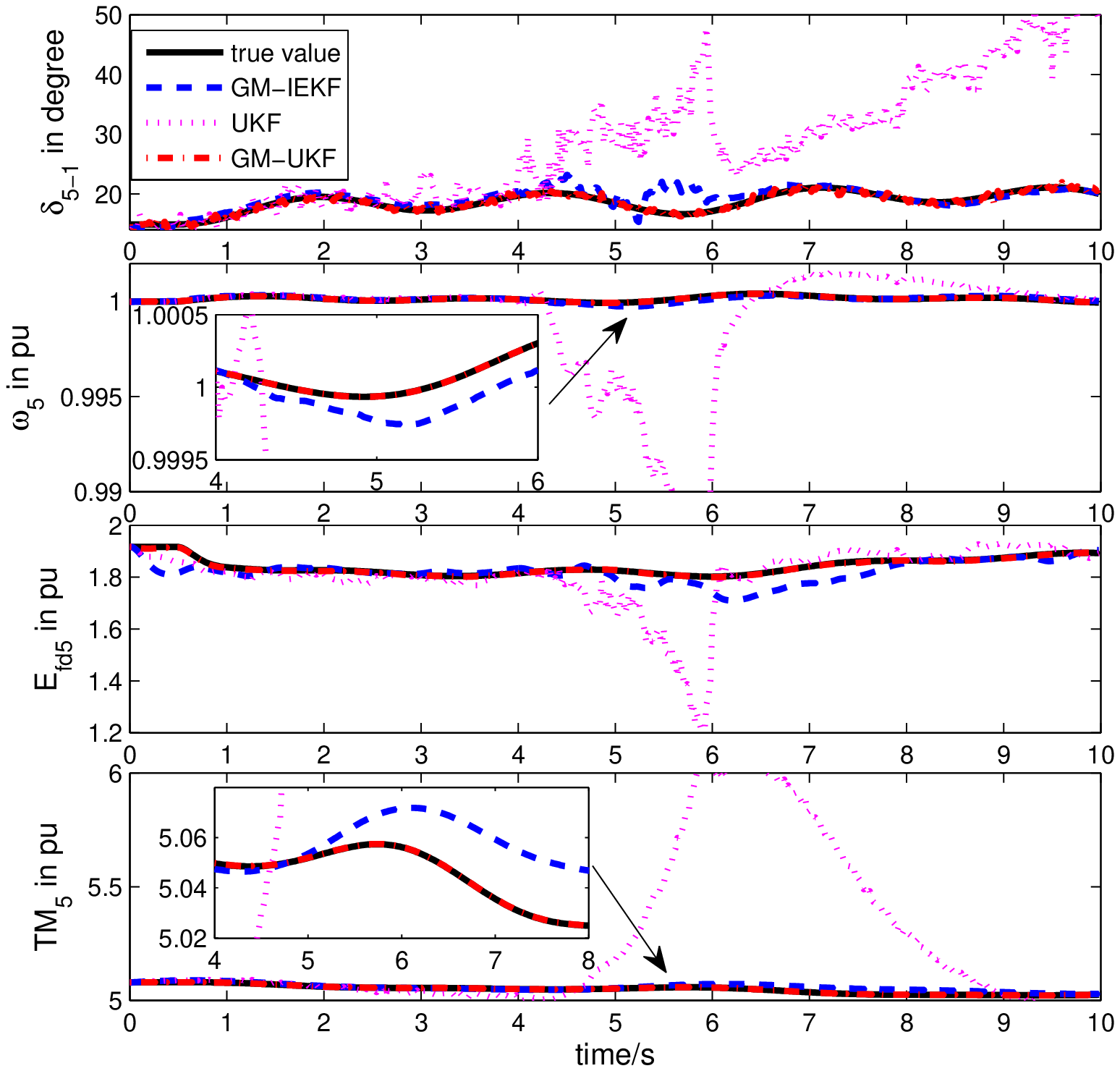}}}
  \mbox{\subfloat[]{\label{subfig:b} \includegraphics[width=7.5cm]{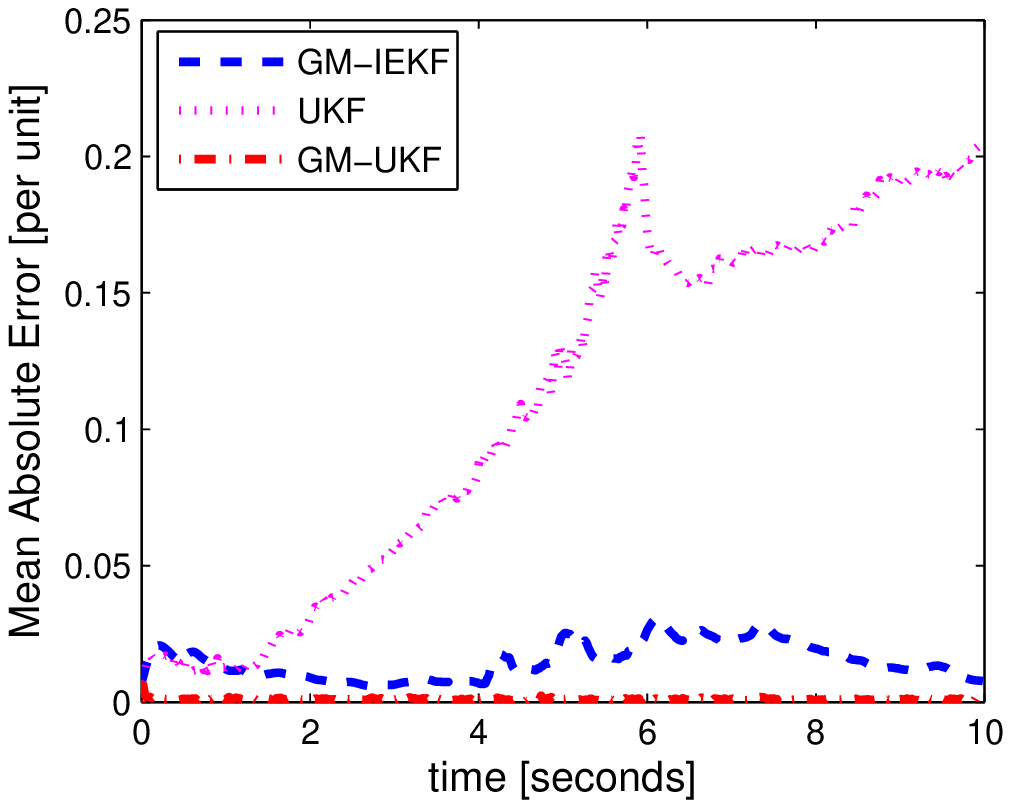}}}
\caption{Case 3: Tracking performance of the GM-IEKF, the UKF, and the GM-UKF in the presence of innovation outliers from $t$=4s to $t$=6s, where the predicted rotor angle of Generator 5 is corrupted with 20\% errors; (a) the estimated rotor angle and speed, field voltage and mechanical power of Generator 5 are used for illustration purposes; (b) mean absolute error of each of the three filters.}
\label{Fig.Laplace_innovation_outliers}
\end{figure}

\begin{figure}
\centering
  \mbox{\subfloat[]{\label{subfig:a} \includegraphics[width=9.5cm]{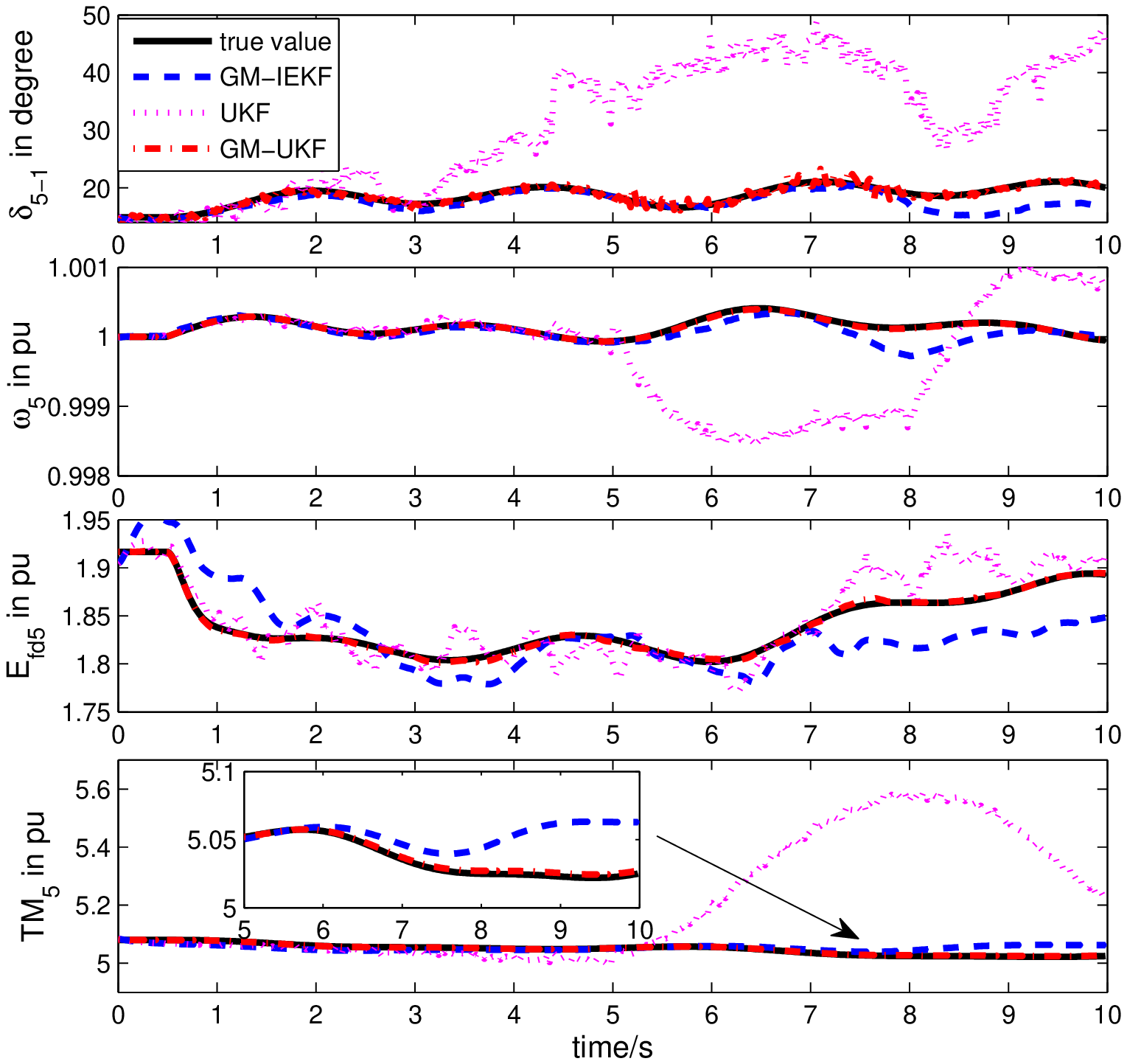}}}
  \mbox{\subfloat[]{\label{subfig:b} \includegraphics[width=7.5cm]{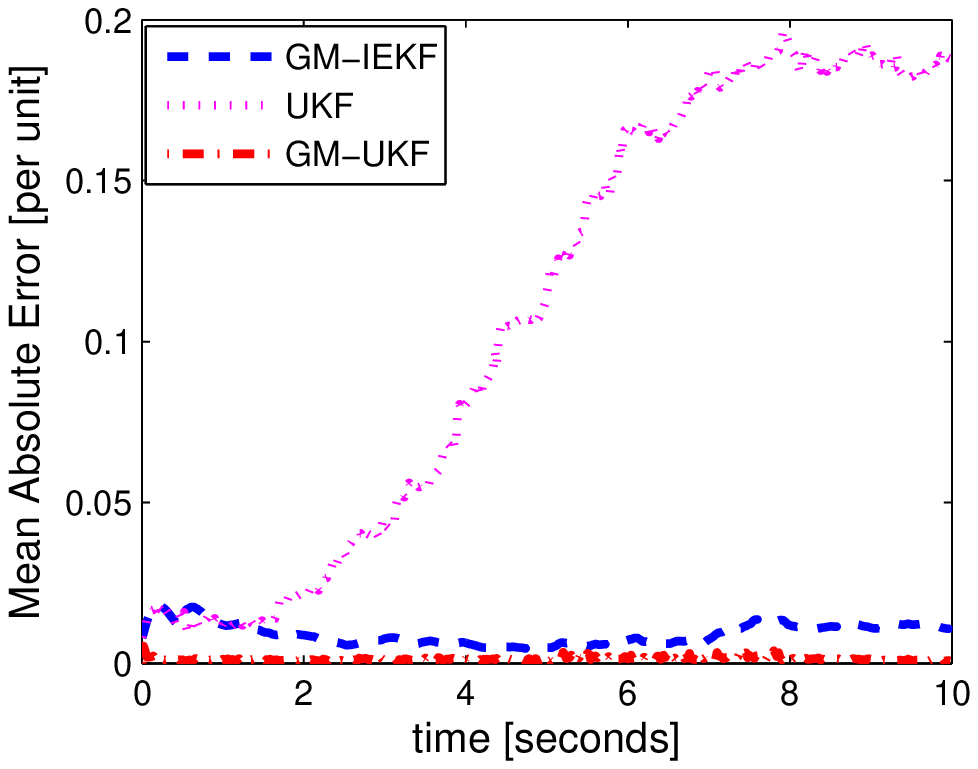}}}
\caption{Case 4: Tracking performance of the GM-IEKF, the UKF, and the GM-UKF in the presence of PMU measurement losses from $t$=5s to $t$=8s, where all the terminal measurements of Generator 5 are lost; (a) the estimated rotor angle and speed, field voltage and mechanical power of Generator 5 are used for illustration purposes; (b) mean absolute error of each of the three filters.}
\label{Fig.Laplace_measurement_loss}
\end{figure}

\begin{figure}
\centering
  \mbox{\subfloat[]{\label{subfig:a} \includegraphics[width=9.5cm]{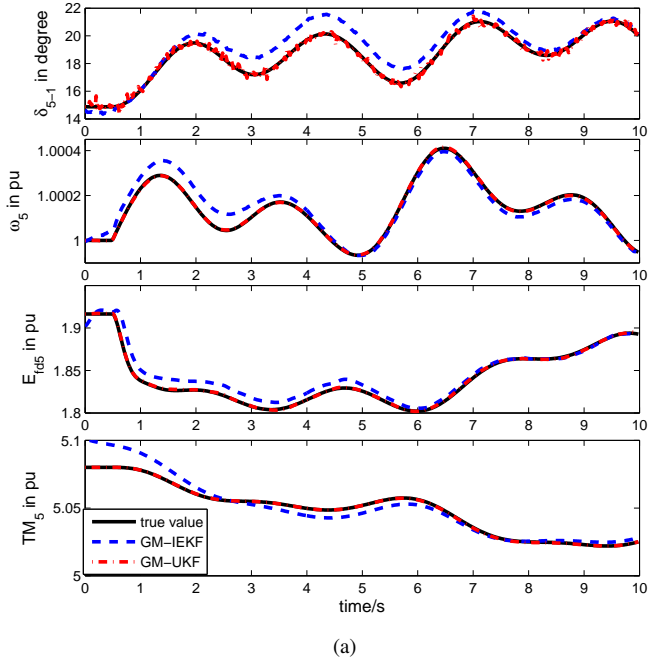}}}
  \mbox{\subfloat[]{\label{subfig:b} \includegraphics[width=7.5cm]{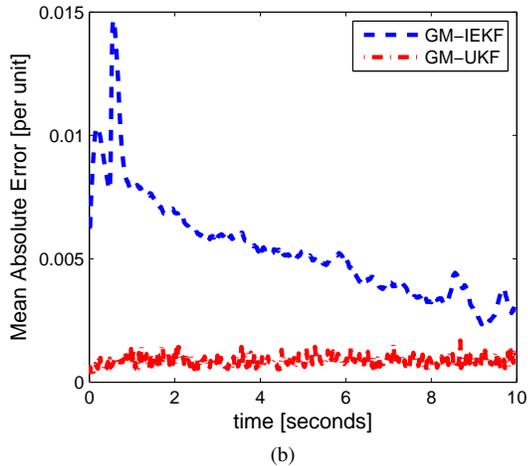}}}
\caption{Case 5: Tracking performance of the GM-IEKF, the UKF, and the GM-UKF in the presence of Cauchy power measurement noise. The estimated rotor angle and speed, field voltage and mechanical power of Generator 5 are used for illustration purposes; (b) mean absolute error of each of the three filters. Since the UKF diverges, its results are not shown in the figure.}
\label{Fig.Cauchy_two_axis_no_outliers}
\end{figure}

\begin{figure}
\centering
  \mbox{\subfloat[]{\label{subfig:a} \includegraphics[width=9.5cm]{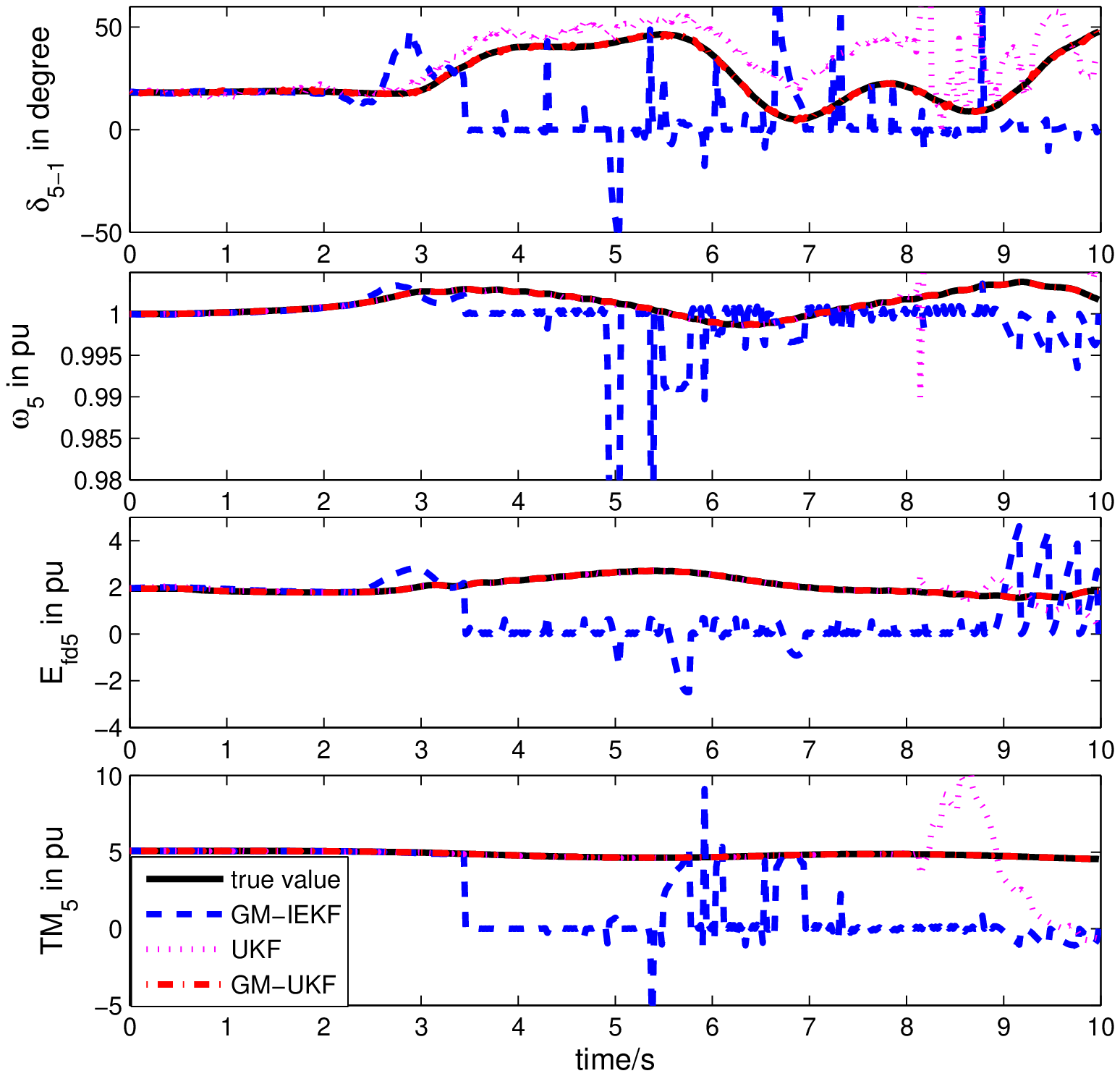}}}
  \mbox{\subfloat[]{\label{subfig:b} \includegraphics[width=8cm]{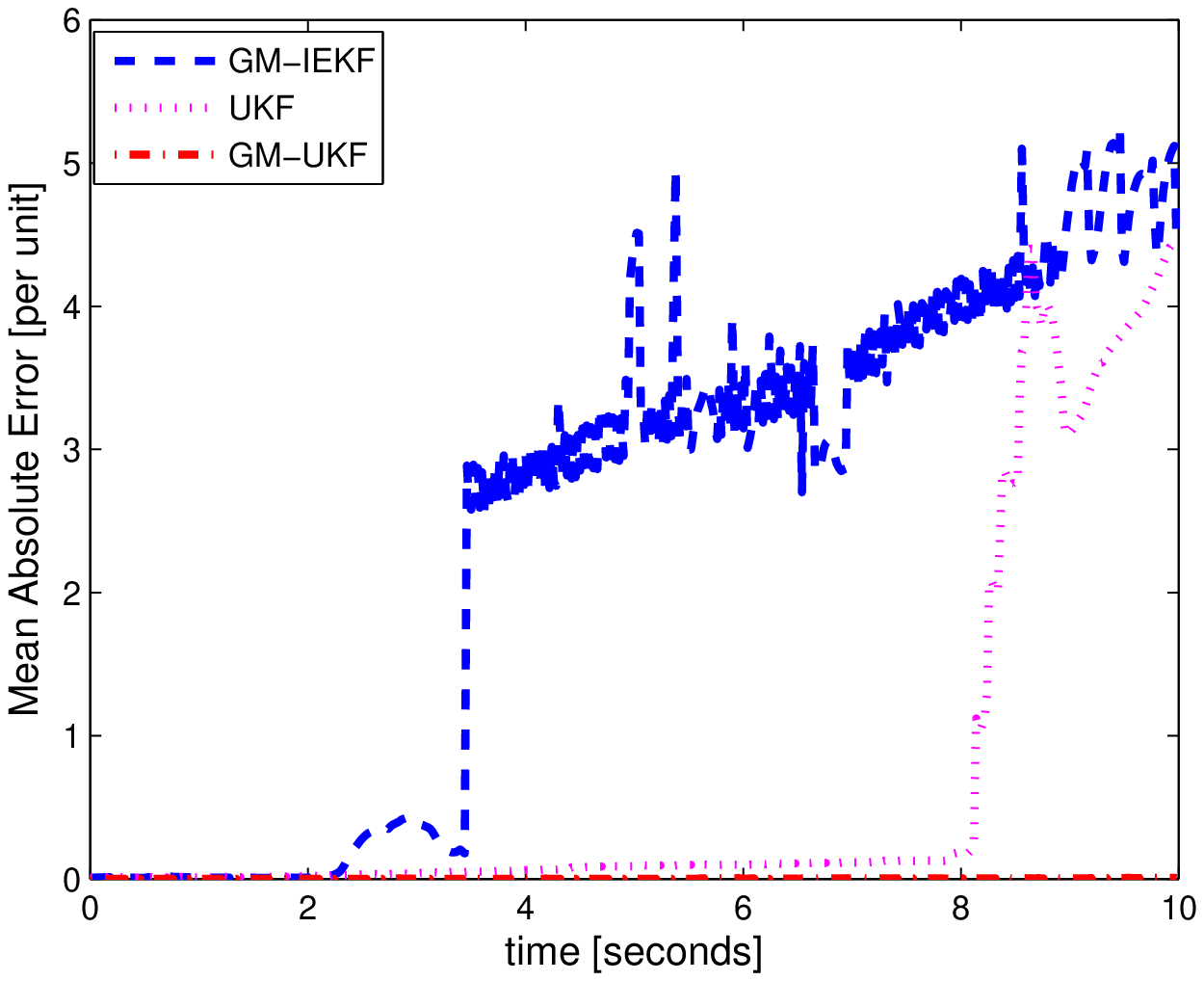}}}
\caption{Tracking performance of the GM-IEKF, the UKF, and the GM-UKF in the presence of strong system nonlinearity. (a) The estimated rotor angle and speed, field voltage and mechanical power of Generator 5 are used for illustration purposes; (b) mean absolute error of each of the three filters.}
\label{Fig.strong_nonlinearity}
\end{figure}

\subsection{Case 2: Thick-tailed Non-Gaussian Measurement Noise with Observation Outliers}
The settings are the same as those of Case 1 except for the presence of observation outliers from $t$=4s to $t$=6s. The latter are simulated by adding 20\% errors to the real and the reactive power measurements of Generator 5. The results are presented in Fig. \ref{Fig.Laplace_observation_outliers}. From this figure, we observe that the UKF is not robust to observation outliers since it yields significantly biased results. Although the GM-IEKF can handle them, it produces increased biases on the estimates at the time when observation outliers occur (see the estimated rotor speed and the field voltage for example). By contrast, the GM-UKF suppresses the outliers and produces much less bias than the GM-IEKF. Note that the Gaussianity of the GM-estimator used in the estimation step of the GM-UKF allows that method to filter out thick-tailed noise while its statistical robustness enables it to suppress the outliers, hence achieving very good estimates.
\vspace{-0.3cm}
\subsection{Case 3: Thick-tailed Non-Gaussian Measurement Noise with Innovation Outliers}
The settings are the same as those of Case 1 except for the presence of innovation outliers from $t$=4s to $t$=6s. They are simulated by adding 20\% errors to the predicted rotor angle of Generator 5. This innovation outlier is induced by a gross parameter value in the model. The comparison results are shown in Fig. \ref{Fig.Laplace_innovation_outliers}. As expected, due to the non-robustness of the UKF, it is unable to handle innovation outliers. On the other hand, the GM-IEKF can handle them, but it produces larger biases compared with Case 2. This can be explained by the fact that the model errors will not only affect the predicted state vector but also will produce a smearing effect throughout the Jacobian matrix. As a result, it downweights several good measurements. By contrast, our GM-UKF is capable of handling both observation and innovation outliers, yielding comparable performances. Because only the sigma points associated with the model errors will be affected and downweighted, this filter obtains better estimates than the GM-IEKF in presence of model errors.
\vspace{-0.5cm}
\subsection{Case 4: Thick-tailed Non-Gaussian Measurement Noise with Measurement Losses}
The settings are the same as those of Case 1 except for the losses of PMU measurement from $t$=5s to $t$=8s; specifically, all the PMU measurements at the terminal bus of Generator 5 are lost due to communication failures or cyber attacks. The test results are presented in Fig. \ref{Fig.Laplace_measurement_loss}. Due to its lack of robustness, the UKF is unable to handle the loss of PMU measurements because in that case, only noise is received and taken as PMU measurements. As for the GM-IEKF and the GM-UKF, thanks to their robustness and to the batch-mode regression form that provides enhanced data redundancy, they will rely on the majority of predicted states and the good measurements to filter out the noise and strongly downweight the lost PMU measurements, which are flagged as outliers by the PS. As a result, they both achieve reasonable state estimates, with an clear advantage for the GM-UKF since it exhibits smaller mean absolute error.
\vspace{-0.3cm}
\subsection{Case 5: Handling Cauchy Power Measurement Noises}
From Part I, it is observed that the real and the reactive power measurement noises may obey a Cauchy distribution, which is a very thick tailed distribution with no moments being defined. To test the capability of our GM-UKF to handle that case, we assume that the simulation settings are the same as those of Case 1 except that the Cauchy noises with zero median and a scale of 0.005 is added to the real and the reactive power injection measurements. The obtained test results are displayed in Fig. \ref{Fig.Cauchy_two_axis_no_outliers}. Note that in presence of Cauchy measurement noise, the UKF has a non-positive definite covariance matrix, resulting in its divergence. By contrast, thanks to the Gaussian normality and robustness of our GM-UKF, the total influence function-based covariance matrix updating approach can always guarantee its positive-definiteness. On the other hand, compared with the results obtained when using Laplacian power measurement noises, the GM-IEKF produces larger biases of the state estimates and takes much longer time to approach the true system states. This is not the case for our GM-UKF since it achieves a comparable performance and tracks the system state at the very beginning of the transient process.
\vspace{-0.3cm}
\subsection{Robustness to Strong System Nonlinearity}
Practical systems may be heavily loaded, resulting in strong nonlinear dynamics. To illustrate the capability of our GM-UKF to handle that case, we assume that the load at Bus 7 is increased from 233.8 MW to 1500 MW to stress the system before switching Line 15-16 while the other simulation settings are the same as those of Case 1. Note that the steady-state maximum loadability at Bus 7 is around 2000 MW. After the line switching, the system operates under even greater stressed conditions. The test results are displayed in Fig. \ref{Fig.strong_nonlinearity}. It is observed from these two figures that the GM-IEKF fails to converge at the very beginning of the transient process while the UKF diverges around $t$=8s. By contrast, our GM-UKF is able to handle this scenario while achieving excellent tracking performance. The underlying reasons are as follows:
\begin{itemize}
  \item Under stressed system operating conditions, the first-order Taylor series expansion used in the EKF and the GM-IEKF is too approximate and is unable to account for strong system nonlinearities. As a result, these two filters produce very large approximation errors and eventually diverge;
  \item Thanks to the sigma-points-based unscented transformation and its approximation accuracy up to at least third-order Taylor series expansion, both the UKF and the GM-UKF are able to handle strong system nonlinearities. However, due to the accumulative estimation error induced by the non-Gaussian measurement error, the estimation error covariance matrix of the UKF is close to non-positive definitiveness. Therefore, it produces large estimation errors and finally diverges. By contrast, our GM-UKF leverages the strength of the unscented transformation to handle system nonlinearities while the robustness of the GM-estimator allows it to filter out thick-tailed non-Gaussian measurement noise, yielding good estimation results.
\end{itemize}
\subsection{Breakdown Point of the GM-UKF to Cyber Attacks}
With the strong reliance of smart grid functions on communication networks, cyber attacks have become a major concern. Typically, they are classified as bias injection attack, denial of service attack, and replay attack \cite{Andre2012,Junbo2016}. Bias injection attack occurs when an adversary attempts to corrupt the content of either the measurement or the control signals; for example, the man-in-the-middle intercepts the PMU measurement signals and corrupts them with large biases. Denial of service attack occurs when the actuator and sensor data are prevented from reaching their respective destinations, resulting in the absence of data for the DSE; for instance, this will be the case if the PMU metered values do not reach the phasor data concentrator. Replay attack occurs when a hacker first performs a disclosure attack from a certain time period, gathering sequences of data, and then begins replaying the data during a certain period; for instance, the current PMU measurements processed by a dynamic state estimator are replaced by past values. In other words, those attacks induce observation or innovation outliers.
\begin{table}
\caption{Average Computing Times of the Three DSE Methods For Every PMU Sample, where NA represents not applicable.}
\label{Tab.computing_time_each_method}
\begin{center}
\normalsize
\begin{tabular}{|c|c|c|c|}
\hline
Cases &UKF& GM-IEKF&GM-UKF\\
\hline
Case 1&6.28ms&9.64ms&9.52ms\\
\hline
Case 2&6.31ms&9.68ms&9.55ms\\
\hline
Case 3&6.38ms&9.72ms&9.63ms\\
\hline
Case 4&6.36ms&9.70ms&9.59ms\\
\hline
Case 5&NA&9.80ms&9.70ms\\
\hline
\end{tabular}
\end{center}
\end{table}

To investigate the breakdown point of the GM-UKF to cyber attacks, which is defined as the maximum number of outliers that the filter can handle without yielding unreliable estimates, we carry out extensive simulations on the IEEE 39-bus test system using the concept of finite sample breakdown in nonlinear regression introduced by Stromberg and Ruppert \cite{Stromberg1992}. By replacing a varying percentage of observations by outliers in the vector $\bm{y}_k$, it is observed that the GM-UKF can handle at least 25\% of corrupted observations. It is worth noting that the breakdown point of the GM-estimator in nonlinear regression is still unknown. This problem will be investigated as a future work. Another interesting problem is the determination of the maximum breakdown point that any regression estimator may have in structured nonlinear regression such as power system state estimation problems; this will be an interesting extension of the results proved in Mili and Coakley \cite{Mili_1996} in the linear case.
\vspace{-0.3cm}
\subsection{Computational Efficiency}
To validate the applicability of the proposed GM-UKF to online estimation with a PMU sampling rate of 30 or 60 samples per second, its computational efficiency is analyzed and compared to that of the UKF and the GM-IEKF in Cases 1-5. The test is performed on a PC with Intel Core i5, 2.50 GHz, 8GB of RAM. The average computing time of each method for every PMU sample is displayed in Table \ref{Tab.computing_time_each_method}. We observe from this table that the UKF has the best computational efficiency, exhibiting computing times much lower than the PMU sampling period, which are 33.3ms and 16.7ms for 30 sample/s and 60 samples/s, respectively. Although the execution times of the GM-IEKF and the GM-UKF is longer, they are still smaller than the PMU sampling period, demonstrating their ability to track system real-time dynamic states.
\vspace{-0.3cm}
\section{Conclusion and Future Work}
In this second part of the two-paper series, the proposed GM-UKF is implemented, tested, and validated. Various scenarios have been considered and explored to evaluate its performance with Laplacian and Cauchy measurement noises, observation and innovation outliers, and strong system nonlinearities. Its breakdown point to cyber attacks has also been investigated. Comparison results with existing methods show that the GM-UKF outperforms the GM-IEKF and the UKF in all the simulated scenarios. It is interesting to note that when the system is operating under stressed conditions, the GM-IEKF and the UKF fail to converge while our GM-UKF converges. Furthermore, if the power measurement noises follow a Cauchy distribution, the UKF fails to converge while the GM-IEKF achieves much lower relative statistical efficiency with respect to our GM-UKF.

There are different possible avenues to further investigate the study considered in this paper. The proposed centralized GM-UKF can handle observation and the innovation outliers, but provides poor results in presence of structural outliers. The latter may be induced by gross errors in circuit breaker statuses or in the parameters of the turbine-generators and the transmission lines. To address this problem, we will investigate a decentralized GM-UKF, which will be implemented at the generating unit level using local voltage and current phasor measurements. Here, additional available measurements on rotor speed, terminal-bus real and reactive power, and field voltage and current can be utilized for improving the measurement redundancy. Furthermore, we will develop a generalized GM-UKF for simultaneously estimating the system states and model parameters whose values are either inaccurate or incorrect. Furthermore, we will investigate the case where the GM-UKF fails to produce good results due to very strong system nonlinearities. The development of the GM-particle filter can be a good candidate to handle that case. Finally, we will extend the proposed GM-UKF to the generator model calibration and validation, the estimation of dynamic load model parameters and power system oscillatory modes.
\vspace{-0.5cm}
\begin{appendices}
\section{Projection Statistics Algorithm}
\label{Implementation_PS}
The main steps of implementing the projection statistics algorithm are shown as follows:
\begin{itemize}
  \item \emph{Step 1}: For a point ${\bm{l}_i}$ in an $n$-dimensional space, calculate the coordinate-wise median given by
\begin{equation}
\bm{M} = \left\{ {\mathop {med}\limits_{j = 1,...,m} \left( {{l_{j1}}} \right),...,\mathop {med}\limits_{j = 1,...,m} \left( {{l_{jn}}} \right)} \right\},
\label{coordiantewise}
\end{equation}
where $m$ is the number of points;
  \item \emph{Step 2}: Calculate the directions for projections ${\bm{u}_j} = {\bm{l}_j} - \bm{M}$, $j = 1,...,m$;
  \item \emph{Step 3}: Normalize ${\bm{u}_j}$ to get
\begin{equation}
{\bm{\ell}_j} = \frac{{{\bm{u}_j}}}{{\left\| {{\bm{u}_j}} \right\|}} = \frac{{{\bm{u}_j}}}{{\sqrt {u_{j1}^2 + ...u_{jn}^2} }};j = 1,...,m;
\label{normalized_directions_PS}
\end{equation}
  \item \emph{Step 4}: Calculate the standardized projections of the vectors $\left\{ {{\bm{l}_1},...,{\bm{l}_m}} \right\}$ on ${\bm{\ell}_j}$, which are given by
\begin{equation}
{{\zeta}_{1j}} = \bm{l}_1^T{\bm{\ell}_j};{{\zeta}_{2j}} = \bm{l}_2^T{\bm{\ell}_j};...,{{\zeta}_{mj}} = \bm{l}_m^T{\bm{\ell}_j};
\label{standardized_P}
\end{equation}
  \item \emph{Step 5}: Calculate the median of $\{{\mathbf{\zeta}_{1j}},...,{{\zeta}_{mj}}\}={{\zeta}_{med,j}}$;
  \item \emph{Step 6}: Calculate the median absolute deviation (MAD) $MAD_j = 1.4826 \cdot {b} \cdot \mathop {med}\limits_i \left| {{{\zeta}_{ij}} - {{\zeta}_{med,j}}} \right|$, where the correction factor is ${b} = 1 + 15/( {m - n})$;
  \item \emph{Step 7}: Calculate the standardized projections
\begin{equation}
{P_{ij}} = \frac{{\left| {{{\zeta}_{ij}} - {{\zeta}_{med,j}}} \right|}}{{MA{D_j}}}{\rm{\; }}for{\rm{\; }}i = 1,...,m
\label{standardized_PS};
\end{equation}
  \item \emph{Step 8}: Repeat steps 4--7 for all vectors $\left\{ {{\bm{\ell}_1},...,{\bm{\ell}_m}} \right\}$ to get the standardized projections $\left\{ {{P_{i1}},...,{P_{im}}} \right\}{\rm{\; }}for{\rm{\; }}i = 1,...,m$;
  \item \emph{Step 9}: Calculate the projection statistics
\begin{equation}
P{S_i} = \max \left\{ {{P_{i1}},...,{P_{im}}} \right\}{\rm{\; }}for{\rm{\; }}i = 1,...,m.
\end{equation}
\end{itemize}
\vspace{-0.3cm}
\section{Proof of the Proposition 1}
\begin{proof}
In the previous companion paper, the estimation error covariance matrix ${\bm{P}_{k\left| k \right.}^{xx}}$ has been shown to follow asymptotic a Gaussian distribution and since the covariance of $\bm{\xi}_{k}$ is an identity matrix, the standardized residual $\bm{r}_S$ therefore follows asymptotic a normal distribution. As a consequence, the probability distribution function of the standardized residual can be expressed as $\phi \left( r_S \right) = \frac{1}{{\sqrt {2\pi } }}{e^{ - \frac{{{r_S{^2}}}}{2}}}$. On the other hand, from the Huber function with $\lambda=1.5$, we can calculate
\begin{equation}
\psi \left( {{r_S}} \right) = \left\{ {\begin{array}{*{20}{c}}
{{r_S}{\rm{ \quad\quad\quad  \quad\quad\;\;          for }}\left| {{r_S}} \right| \le \lambda {\rm{ }}}\\
{\lambda sign\left( {{r_S}} \right){\rm{\quad\quad for }}\left| {{r_S}} \right| > \lambda }
\end{array}} \right.,
\label{Eq:Huberfunctionfirstorderderivative}
\end{equation}
\begin{equation}
{\psi ^{'}}\left( r_S \right) = \left\{ \begin{array}{l}
1{\rm{ \quad\quad\quad    for }}\left| r_S \right| \le \lambda \\
0{\rm{ \quad\quad\quad    for }}\left| r_S \right| > \lambda {\rm{ }}
\end{array} \right..
\label{Eq:Huberfunctionsecondorderderivative}
\end{equation}
Then, we can further obtain
\begin{IEEEeqnarray}{lll}
\mathbb{E}\left[ {{\psi ^{'}}\left( r_S \right)} \right] = \int_{ - \infty }^\infty  {{\psi ^{'}}\left( r_S \right)\phi \left( r_S \right)dr_S = } \frac{1}{{\sqrt {2\pi } }}\int_{ - \infty }^\infty  {{e^{ - \frac{{{r_S^2}}}{2}}}dr_S} \nonumber\\
{\rm{ \quad\quad\quad\quad\;\;  }} = 2\Phi \left( \lambda  \right) - 1 = 0.8664,
\label{Eq:expectationoffirstorder}
\end{IEEEeqnarray}
\begin{IEEEeqnarray}{lll}
\mathbb{E}\left[ {{\psi ^2}\left( r_S \right)} \right] = \int_{ - \infty }^\infty  {{\psi ^2}\left( r_S \right)\phi \left( r_S \right)dr_S} \nonumber\\
 = \frac{{{\lambda ^2}}}{{\sqrt {2\pi } }}\int_{ - \infty }^{ - \lambda } {{e^{ - \frac{{{r_S^2}}}{2}}}dr_S}  + \frac{1}{{\sqrt {2\pi } }}\int_{ - \lambda }^\lambda  {{r_S^2}{e^{ - \frac{{{r_S^2}}}{2}}}dr_S}  \nonumber\\
 \quad\quad\quad\quad\quad\quad\quad\quad\quad\quad\quad\quad\quad + \frac{{{\lambda ^2}}}{{\sqrt {2\pi } }}\int_b^\infty  {{e^{ - \frac{{{r_S^2}}}{2}}}dr_S} \nonumber\\
 = {\lambda ^2}\Phi \left( { - \lambda } \right) - \frac{{2\lambda }}{{\sqrt {2\pi } }}{e^{ - \frac{{{\lambda ^2}}}{2}}} + 2\Phi \left( \lambda  \right) - 1 + {\lambda ^2}\left( {1 - \Phi \left( \lambda  \right)} \right)\nonumber\\
 = 0.7784.
\label{Eq:expectationoforder}
\end{IEEEeqnarray}
Finally, we can calculate
\begin{equation}
\frac{{\mathbb{E}\left[ {{\psi ^2}\left( r_S \right)} \right]}}{{{{\left( {\mathbb{E}\left[ {{\psi ^{'}}\left( r_S \right)} \right]} \right)}^2}}} = \frac{{0.7784}}{{{{\left( {0.8664} \right)}^2}}} = 1.0369.
\label{Eq:expectationresult}
\end{equation}
\end{proof}
\end{appendices}
\vspace{-0.3cm}
\section*{Acknowledgment}
The authors would like to thank Dr. Zhenyu Huang from PNNL for providing us with real PMU data for analyzing the
statistical probability distributions of the PMU measurement errors.

\ifCLASSOPTIONcaptionsoff
  \newpage
\fi
\end{document}